\documentclass[10pt,journal,twocolumn]{IEEEtran}
\IEEEoverridecommandlockouts
\usepackage{epsfig,cite,bm,algorithm,algorithmic,epstopdf,amsmath,subfigure,multirow,amssymb,amsfonts,amsthm,xcolor,caption,graphicx,textcomp,xcolor, subfigure, makecell, bbding, soul}
\usepackage[inline, shortlabels]{enumitem}
\usepackage[flushleft]{threeparttable}
\soulregister{\cite}7
\soulregister{\ref}7
\setlist{itemjoin ={,\enspace},itemjoin* = { and\enspace}}
\def\QEDclosed{\mbox{\rule[0pt]{1.5ex}{1.5ex}}}
\usepackage[T1]{fontenc}
\usepackage{stfloats}
\allowdisplaybreaks[4]

\allowdisplaybreaks[4]
\usepackage{booktabs}

\begin{document}

\title{Sub-Connected Hybrid Beamfocusing Design for RSMA-Enabled Near-Field Communications with Imperfect CSI and SIC}
	
\author{Jiasi Zhou, Ruirui Chen, Yanjing Sun,~\IEEEmembership{Member,~IEEE}, and Chintha Tellambura,~\IEEEmembership{Fellow,~IEEE}
\thanks{Jiasi Zhou is with the School of Medical Information and Engineering, Xuzhou Medical University, Xuzhou, 221004, China, (email: jiasi\_zhou@xzhmu.edu.cn). (\emph{Corresponding author: Jiasi Zhou}).}
\thanks{Ruirui Chen and Yanjing Sun are with the School of Information and Control Engineering, China University of Mining and Technology, Xuzhou 221116, China (email: \{rrchen, yjsun\}@cumt.edu.cn).}
\thanks{ Chintha Tellambura is with the Department of Electrical and Computer Engineering, University of Alberta, Edmonton, AB, T6G 2R3, Canada (email: ct4@ualberta.ca).} 
\thanks{This work was supported by the Talented Scientific Research Foundation of Xuzhou Medical University (D2022027) and National Natural Science Foundation of China (62441115).}}
\maketitle

\begin{abstract}
Near-field spherical waves inherently encode both direction and distance, enabling spotlight-like beamfocusing for targeted interference mitigation. However, whether beamfocusing alone can fully suppress interference under perfect or imperfect channel state information (CSI), thereby eliminating the need for advanced interference management, remains unclear. To address this, we investigate rate-splitting multiple access (RSMA)-enabled near-field communications (NFC) under imperfect CSI. Our transmit scheme employs a sub-connected hybrid analog-digital (HAD) architecture to reduce hardware complexity and incorporates imperfect successive interference cancellation (SIC) for practical deployment. We formulate a max-min rate optimization problem that jointly designs the analog beamfocuser, digital beamfocuser, and common rate allocation. A penalty-based block coordinate descent (BCD) algorithm is proposed to solve the non-convex problem, with closed-form solutions derived for the optimal beamfocusers. To further reduce complexity, a low-complexity two-stage algorithm is introduced, where analog and digital beamfocusers are designed separately. Simulation results show that: (1) beamfocusing alone cannot fully mitigate interference, even with perfect CSI; (2) RSMA offers superior interference management compared to SDMA under imperfect CSI and SIC; and (3) the sub-connected HAD architecture achieves near-optimal performance with significantly fewer RF chains.
\end{abstract} 

\begin{IEEEkeywords}
Near-field communications, rate splitting multiple access, imperfect channel state information, imperfect successive interference cancellation.
\end{IEEEkeywords}

\section{Introduction} 
The exponential growth of wireless devices has revealed fundamental limitations in conventional orthogonal multiple access (OMA) schemes \cite{mao2018rate}. To support massive connectivity and high data rates, modern systems increasingly adopt spatial division multiple access (SDMA) and non-orthogonal multiple access (NOMA) \cite{10630543}. However, SDMA treats interference as noise, which leads to performance saturation when interference dominates the signal \cite{9835151}. On the other hand, NOMA requires decoding all stronger interference signals, demanding complex receiver designs. As a result, both approaches exhibit limited flexibility in interference management, ultimately constraining network efficiency \cite{mao2018rate,10273395}.

To address these challenges, rate-splitting multiple access (RSMA) has emerged as a more flexible and robust alternative \cite{10038476}. RSMA enables users to partially decode interference while treating the remaining part as noise. This tunable interference decoding capability allows dynamic adaptation to varying interference levels, effectively unifying and generalizing SDMA and NOMA as exceptional cases. Consequently, RSMA provides a more versatile transmission strategy, positioning itself as a promising multiple access framework for sixth-generation (6G) wireless networks.

RSMA is known to outperform SDMA and NOMA in far-field communications by achieving higher transmission rates and greater robustness \cite{10159012,loli2022rate,10721401}. However, emerging applications such as autonomous vehicles are pushing wireless networks toward higher frequency bands and the use of extremely large-scale antenna arrays (ELAAs) \cite{10558818}. These developments extend the Rayleigh distance and shift wave propagation from far-field planar to near-field spherical characteristics. Unlike planar waves, spherical waves embed both direction and distance information, enabling dual-dimensional resolution. This allows for spotlight-like beamfocusing that concentrates energy on specific spatial locations, enabling precise signal enhancement and targeted interference mitigation \cite{10663521}.

Given their respective advantages, RSMA-enabled near-field beamfocusing has been preliminarily explored \cite{10988830,10798456,10906379,11071287,10414053,zhou2024hybrid,zhou2025crb}, demonstrating notable performance gains. However, these pioneering works primarily address complex interference scenarios, such as the coexistence of communication and sensing, where RSMA's role becomes essential. In contrast, for near-field communication (NFC) scenarios involving only multi-user interference (MUI), two fundamental questions remain open, as summarized below.

\begin{enumerate}
\item \textbf{Can near-field beamfocusing achieve ideal MUI suppression under perfect and imperfect channel state information (CSI)? } This question is crucial, as CSI acquisition is particularly challenging in NFC systems due to the use of ELAAs \cite{10496996}. Unlike far-field beams, near-field beams can precisely concentrate energy on specific spatial locations with minimal leakage to neighboring regions. This spatial resolution allows for effective user separation based on location, inherently eliminating  MUI \cite{10135096,10559261}. This distinctive property raises the possibility that NFC systems might not require advanced interference management schemes. However, it remains unclear whether imperfect CSI undermines this interference suppression capability, thereby necessitating complementary strategies -- such as RSMA -- for achieving robust and optimal performance.

\item \textbf{Does RSMA maintain its performance advantage in NFC under imperfect successive interference cancellation (SIC)?} SIC is fundamental to the operation of RSMA, which operates as follows. The base station divides each user’s message into a common part and a private part, combining all common parts into a single common stream and encoding each private part into a separate private stream. At the receiver, each user decodes the common stream, cancels it via SIC, and then retrieves its private stream. When SIC is imperfect, residual interference from the common stream degrades the signal quality of the private streams, reducing the overall transmit rate. This compromises RSMA’s ability to achieve its theoretical rate gains and weakens its robustness to CSI imperfections. While RSMA has demonstrated strong performance in NFC systems under ideal SIC conditions, practical implementations inevitably face SIC imperfections \cite{10028754}, which may limit its effectiveness. This raises a key question: Can RSMA still deliver meaningful performance gains in NFC despite imperfect SIC? 
\end{enumerate}

Addressing these two issues can provide crucial guidelines for designing NFC transmission frameworks. However, to our knowledge, these two problems remain unaddressed, thereby forming the primary motivation for this work.

\subsection{Related Works}
\subsubsection{SDMA/NOMA-enabled NFC} The potential benefits of near-field beamfocusing has been widely explored in\cite{9738442,10123941,10436390,10579914,zuo2023non}. For example, the feasibility of near-field beamfocusing is investigated across three different antenna architectures: full digital, hybrid analog-digital (HAD), and dynamic metasurface antennas\cite{9738442}. The study reveals that the distance-dependent characteristic of spherical wave channels facilitates precise signal focusing at target locations, effectively suppressing MUI. A key theoretical contribution is established in \cite{10123941}, where the asymptotic orthogonality of near-field beamfocusing vectors in the distance domain is rigorously proven. Leveraging this finding, the authors propose a novel location division multiple access (LDMA) scheme, opening up new possibilities for improving network performance\cite{10123941}. The intrinsic beamfocusing capability can safeguard against information leakage without requiring additional countermeasures, even when eavesdroppers share the same angular alignment as legitimate users\cite{10436390}. Furthermore, the additional distance dimension in spherical wave propagation enables joint direction and distance estimation in integrated sensing and communications (ISAC) with limited bandwidth\cite{10579914}. The aforementioned publications primarily employ SDMA while leveraging spatial beamfocusing capability to counter MUI\cite{9738442,10123941,10436390,10579914}. To further enhance connectivity, the authors in \cite{zuo2023non} propose a NOMA-based transmit scheme for NFC, which implements a far-to-near SIC decoding order. Their findings highlight the crucial transition to NFC paradigms.

\subsubsection{RSMA-enabled NFC}
To mitigate MUI in   NFC and near-field ISAC systems, RSMA-based transmit schemes have been developed \cite{10988830,10798456,10906379,11071287,10414053,zhou2024hybrid,zhou2025crb}. For example, near-field beamfocusing vectors at both base stations and RIS are jointly optimized to enhance network throughput \cite{10988830}. In RSMA-enabled NFC, preconfigured spatial beams are reused to support additional users \cite{10798456} or detect near-field targets \cite{10906379}. Reference  \cite{11071287} considers NFC with imperfect CSI and SIC, but adopts a mixed propagation model alternating between near-field spherical waves and far-field Rayleigh fading. This leads to a complex hybrid scenario, necessitating sophisticated and advanced interference management strategies. Notably, these approaches rely on fully digital antenna architectures, which require one RF chain per antenna and are impractical for NFC systems with ELAAs \cite{10220205}. To reduce hardware complexity, reference\cite{10414053} proposes a fully-connected HAD architecture for RSMA-enabled mixed-field communication. Similarly, RSMA-based schemes for near-field ISAC under HAD architectures evaluate sensing performance using detection rate \cite{zhou2024hybrid} and CRB \cite{zhou2025crb}. While these works address complex interference scenarios, the two fundamental questions raised earlier remain unexplored.

\subsection{Motivations and Contributions}
To address these research gaps, we investigate an RSMA-enabled NFC system under imperfect CSI and SIC conditions. A sub-connected HAD architecture is employed to achieve hardware efficiency while maintaining system performance. Our principal contributions are as follows:

\begin{itemize}
\item \textbf{RSMA-based NFC Scheme:} We exploit the sub-connected HAD architecture to develop this, specifically designed to address the practical challenges of imperfect CSI and SIC. RSMA provides flexible MUI management while the sub-connected HAD architecture alleviates hardware overhead. The primary objective is to maximize the minimum rate through the joint optimization of analog beamfocuser, digital beamfocuser, and common rate allocation variables.

\item \textbf{Algorithm Design:} We recast the objective function and constraints into tractable forms by adopting a penalty-based approach and introducing auxiliary variables. Subsequently, we partition all variables into three blocks and employ coordinate descent (BCD) methodology to optimize them in an alternating manner.
\begin{enumerate}
\item \emph{Auxiliary variables and common rate allocation optimization}: To attack this subproblem, we leverage surrogate optimization, constructing concave quadratic surrogates to approximate complex logarithmic transmission rates. These constructed surrogates meet the minorization property and gradient consistency, ensuring that the proposed iterative algorithm achieves superior convergence performance\cite{mairal2013optimization}.
\item \emph{Analog beamfocuser optimization:} The optimal analog beamfocuser is derived in closed form. 
\item \emph{Digital beamfocuser optimization:} Similarly, the optimal digital beamfocuser is obtained via a closed-form solution.  
\end{enumerate}

\item \textbf{Low-complexity Algorithm:} To reduce computational complexity, we propose a low-complexity two-stage beamfocusing design algorithm. Specifically, the analog beamfocusing stage maximizes the minimum array gain across all users, while the digital beamfocusing stage optimizes the minimum achievable rate via the low-dimensional equivalent channel. The proposed low-complexity algorithm avoids the double-loop iteration.

\item \textbf{Numerical Insights:} Numerical results demonstrate the effectiveness of the RSMA-based transmit scheme, highlighting three key advantages: 1) superior interference management under imperfect CSI and SIC conditions; 2) considerable performance gains compared to conventional near-field beamfocusing and far-field beamforming; 3) near-optimal digital beamfocusing performance with fewer RF chains. 
\end{itemize}

\emph{Organization:}  The remainder of this paper is organized as follows.  Section \ref{Section II} introduces the signal model and formulates a max-min rate optimization problem. Section \ref{Section III} develops a penalty-based iterative algorithm. Section \ref{Section IV} presents the proposed low-complexity algorithm. Section \ref{Section V} provides the numerical results. Finally, Section \ref{Section VI} concludes this paper.

\emph{Notations:} Boldface upper-case letters, boldface lower-case letters, and calligraphy letters denote matrices, vectors, and sets. The complex matrix space of $N\times K$ dimensions is denoted by $\mathbb{C}^{N\times K}$. The superscripts ${(\bullet)}^T$ and ${(\bullet)}^H$ represent the transpose and Hermitian transpose, respectively. $\text{Re}\left( \bullet\right)$, $\text{Tr}\left( \bullet\right)$, and $\mathbb{E}\left[\bullet\right]$ 
 denote the real part, trace, and statistical expectation. $\text{diag}\left( \bullet\right)$  and $\text{blkdiag}\left( \bullet\right)$ denote diagonal and block diagonal operations, respectively. Operator $\lfloor a\rfloor$  is the largest integer not greater than $a$. The Frobenius norm of matrix $\mathbf{X}$ is denoted by $||\mathbf{X}||_F$. For matrix $\mathbf{X}$, $\mathbf{X}\left(i:j,:\right)$  represent a sub-matrix composed of the rows from the $i$-th to $j$-th. For vector $\mathbf{x}_i$, $\mathbf{x}_{i,j}$  represent its $j$-th element. Variable  $x\sim\mathcal{CN}(\mu, \sigma^2)$ is a  circularly symmetric complex Gaussian (CSCG) with mean $\mu$ and variance $\sigma^2$.

\section{System Model and Problem Formulation}\label{Section II}
This paper considers an RSMA-enabled NFC system, where a BS equipped with $N$ antennas and $L$ RF chains serves $K$ single-antenna users. The sets of transmit antennas, RF chains, and users are indexed as $\mathcal{N}=\left\{1,\dots,N\right\}$, $\mathcal{L}=\left\{1,\dots,L\right\}$, and $\mathcal{K}=\left\{1,\dots,K\right\}$, respectively. The BS employs an uniform linear array (ULA) with an inter-antenna of $d$, yielding a Rayleigh distance of $d_{\text{r}}=\frac{2D^2}{\lambda}$, where $D=\left(N-1\right)d$ and $\lambda$ are the array aperture and carrier frequency, respectively.

\subsection{Near-field channel model with imperfect CSI}
Compared to far-field CSI, near-field CSI acquisition presents significantly greater challenges in achieving high accuracy. To account for the practical constraint, we consider imperfect CSI conditions and characterize the uncertainty using a norm-bounded error model. Formally,  the actual channel vector $\mathbf{h}_k\in\mathbb{C}^{N\times 1}$ between the BS and the $k$-th user is modeled as
\begin{align}
\mathbf{h}_k = \hat{\mathbf{h}}_k+\tilde{\mathbf{h}}_k.
\label{Imperfect_Channel}
\end{align}
where $\hat{\mathbf{h}}_k\in\mathbb{C}^{N\times 1}$ is the estimated CSI, $\tilde{\mathbf{h}}_k\in\mathbb{C}^{N\times 1}$ represents the estimation error, and $||\tilde{\mathbf{h}}_k||_2\leq\epsilon_k$ defines the uncertainty region with $\epsilon_k$ specifying the maximum error norm.

We next elaborate on the estimated channel vector $\hat{\mathbf{h}}_k$ based on spherical wavefronts. Without loss of generality, we define the midpoint of the ULA at the origin with the array aligned with the $y$-axis. In this setting, the coordinate of the $n$-th antenna is $\mathbf{s}_n=\left(0, \tilde nd\right)$, where $n\in\mathcal{N}$ and $\tilde n=\frac{2n-N-1}{2}$. Let $d_k$ and $\theta_k$ denote the distance and angle of user $k$ to the origin of the coordinate system, so its coordinate is $\mathbf{r}_k=\left(r_k\cos\theta_k,r_k\sin\theta_k\right)$. Therefore, the distance between the $n$-th antenna and the $k$-th user is 
\begin{align}
d_{k,n}=||\mathbf{r}_k-\mathbf{s}_n|| = \sqrt{r^2_k+(\tilde nd)^2-2\tilde ndr_k\sin\theta_k}.
\end{align}
Moreover, the channel gain across all links is approximately uniform in the Fresnel region ($1.2D\leq r_k\leq d_{\text{r}}$)\cite{10579914}. Consequently, the channel gain for each link can be approximated by the free-space pathloss of the central link, expressed as $\tilde\beta_k=\frac{c}{4\pi fr_k}$, where $f$ and $c$ are the carrier frequency and speed of light, respectively. As such, the channel between the $n$-th antenna and the $k$-th user can be modeled as
\begin{align}
 \hat h_{k,n}= \tilde\beta_{k} e^{-j\frac{2\pi}{\lambda}d_{k,n}}= \beta_{k} e^{-j\frac{2\pi}{\lambda}\left(d_{k,n}-r_k\right)},
\label{Single_Channel}
\end{align}
where $\beta_{k,n}=\tilde\beta_{k,n}e^{-j\frac{2\pi}{\lambda}r_k}$ denotes the complex channel gain. Using the second-order Taylor expansion to $d_{k,n}$, the channel $\hat h_{k,n}$ can be approximated as $\hat h_{k,n}\approx \beta_{k} e^{j\frac{2\pi}{\lambda}\delta_{k,n}}$, where $\delta_{k,n}=\tilde nd\sin\theta_k-(\tilde nd)^2\cos^2\theta_k/2r_k$. Aggregating all antenna elements, the near-field channel vector $\hat{\mathbf{h}}_k$ between the BS and the $k$-user is given by
\begin{align}
\hat{\mathbf{h}}_k = \beta_{k}\left[e^{j\frac{2\pi}{\lambda}\delta_{k,1}},\dots,e^{j\frac{2\pi}{\lambda}\delta_{k,N}}\right]^T=\beta_{k}\mathbf{a}\left(r_k,\theta_k\right).
\label{Channel}
\end{align}
where $\mathbf{a}\left(r_k,\theta_k\right)$ denotes the near-field array response vector.

\subsection{RSMA-enabled signal encoding}
\begin{figure*}[tbp]
\centering
\includegraphics[scale=0.65]{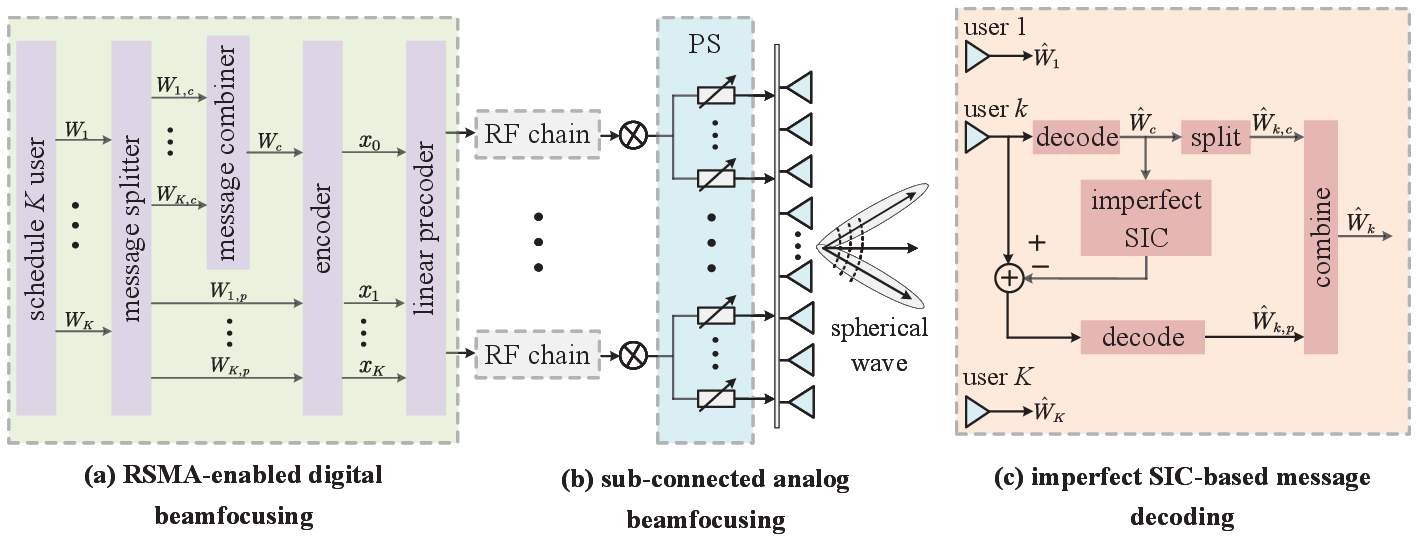}
\caption{The considered RSMA-enabled NFC under imperfect CSI and SIC conditions.}
\label{system}
\end{figure*}
Based on RSMA design criteria, as Fig.~\ref{system}(a), the BS splits the intended message for the $k$-th user into a common part $W_{k,c}$ and a private part $W_{k,p}$. All common parts $\left\{W_{1,c},\dots,W_{K,c}\right\}$ are jointly encoded into a common stream $x_0$ expected to be received and decoded by all users. Concurrently, the private parts $\left\{W_{1,p},\dots,W_{K,p}\right\}$ are independently encoded into $K$ user-specific private streams $\left\{x_{1},\dots,x_{K}\right\}$. All streams maintain mutual independence and unit power normalization, i.e.,  $\mathbb{E}|x_{k}|^2=1$ for $\forall k \in\tilde{\mathcal{K}}=\{0,1,\dots, K\}$. These streams are then linearly precoded and superimposed by HAD beamfocuser $\mathbf{FW}\in\mathbb{C}^{N\times (K+1)}$, where $\mathbf{F}\in\mathbb{C}^{N\times L}$ and $\mathbf{W}=\left[\mathbf{w}_0,\mathbf{w}_1,\dots, \mathbf{w}_K\right]\in\mathbb{C}^{L\times (K+1)}$ are respectively the analog beamfocuser and baseband digital beamfocuser. Here, $\mathbf{w}_0\in\mathbb{C}^{L\times 1}$ and $\mathbf{w}_k\in\mathbb{C}^{L\times 1}$ correspond to the beamfocusing vectors for the common stream and the $k$-th private stream, respectively. The resultant transmit signal can be given by $\mathbf{x}=\mathbf{Fw}_0x_0+\sum^{K}_{k=1}\mathbf{Fw}_kx_k$.

To reduce hardware cost, we employ a sub-connected phase-shifted analog beamfocusing architecture, as Fig.~\ref{system}(b). Specifically, this configuration interfaces between $L$ RF chains and the transmit antenna array. Each RF chain drives a distinct sub-array of $M = N/L$ antennas. As a result, the corresponding analog beamfocuser $\mathbf{F}$ has a block-diagonal structure, expressed as
\begin{equation}
\mathbf{F}=\text{blkdiag}\left(\mathbf{f}_1,\dots,\mathbf{f}_{L}\right)\in\mathbb{C}^{N\times L},
\end{equation}
where $\mathbf{f}_{l}\in\mathbb{C}^{M\times 1}$ contains the phase-shift (PS) coefficients of the $l$-th RF chain. However, due to the inherent hardware limitation, all non-zero elements must satisfy the unit-modulus constraint, i.e., $\left|\mathbf{f}_{l,m}\right|=1$ for $l\in\mathcal{L}$ and $m\in\mathcal{M}=\{1,\dots,M\}$.

\subsection{Imperfect SIC-based signal decoding}
The received signal at user $k$ for $\forall k\in\mathcal{K}$ is given by
\begin{equation}
y_{k}=\underbrace{\mathbf{h}^H_k\mathbf{F}\mathbf{w}_{0}x_{0}}_{\text{Common signal }y_{k,c}}+ \underbrace{\sum\nolimits^K_{i=1}\mathbf{h}^H_k\mathbf{F}\mathbf{w}_{i}x_{i}}_{\text{Private signal }y_{k,p}}+\underbrace{n_{k}}_{\text{noise}},
\end{equation}
where $n_{k}\sim \mathcal{CN}\left(0,\sigma^2_{k}\right)$ denotes additional white Gaussian noise (AWGN) term. With the RSMA decoding principle, as Fig.~\ref{system}(c), each user decodes the common stream by treating all private streams as noise, so the corresponding  signal-to-interference-plus-noise ratio (SINR) is given by
\begin{equation}
\gamma_{k,c}=\frac{\left|\mathbf{h}^H_k\mathbf{Fw}_0\right|^2}{\sum^{K}_{i=1}\left|\mathbf{h}^H_k\mathbf{Fw}_i\right|^2 + \sigma^2_k}.
\end{equation}
We adopt the average transmit rate in terms of the CSI estimation error as the communication metric, which is given by 
\begin{equation}
R_{k,c}=\mathbb{E}_{\tilde{\mathbf{h}}_{k}}\Big(\log\left(1+\gamma_{k,c}\right)\Big).
\label{Common_rate}
\end{equation}
To ensure that all users can successfully decode the common stream, the common rate shall not exceed $R_c= \min_{\forall k}R_{k,c}$. Moreover, since all users share the common rate,  we have $R_c=\sum_{k=1}^{K}C_{k,c}$, where $C_{k,c}$ is the portion of the common rate transmitting $W_{k,c}$. 

Subsequently, the $k$-th user reconstructs the common stream and performs SIC to remove it from the received signal. Under ideal conditions with perfect SIC implementation, the residual signal can be expressed as $y^{\text{SIC}}_{k}=y_{k,p}+n_k$. However, in practice, hardware impairments inevitably lead to incomplete interference cancellation.  As a result, the post-SIC signal can be expressed as $y^{\text{iSIC}} = \Delta_k y_{k,c}+y_{k,p}+n_k$, where $\Delta_k\in[0,1]$ quantifies the residual common signal percentage after imperfect cancellation, i.e., $\Delta_k=0$ denotes perfect SIC. As a result, the SINR of decoding the desired private stream is given by
\begin{equation}
\gamma_{k,p}=\frac{\left|\mathbf{h}^H_k\mathbf{Fw}_k\right|^2}{\Delta_k\left|\mathbf{h}^H_k\mathbf{Fw}_0\right|^2+\sum^{K}_{i=1,i\neq k}\left|\mathbf{h}^H_k\mathbf{Fw}_i\right|^2 + \sigma^2_k}.
\end{equation}
The transmit rate of the $k$-th user's private stream is 
\begin{equation}
R_{k,p}=\mathbb{E}_{\tilde{\mathbf{h}}_{k}}\Big(\log\left(1+\gamma_{k,p}\right)\Big).
\label{Private_rate}
\end{equation}
Consequently, the total transmit rate for the $k$-th user is $R_k=C_{k,c}+R_{k,p}$.

\subsection{Problem formulation}
This paper aims to maximize the minimum rate among all users by jointly optimizing the analog beamfocuser, digital beamfocuser, and common rate allocation. Therefore, the problem is formulated as
\begin{subequations}\label{linear_p}
	\begin{align}
&\max_{\mathbf{F},\mathbf{W},\mathbf{c} } \min_{\forall k} R_k,\label{ob_a}\\
	\text{s.t.}~
	&||\mathbf{FW}||^2_F\leq P_{\text{th}},\label{ob_b}\\
 &\sum_{k=1}^{K}C_{k,c} \leq R_{c},\label{ob_c}\\
 &C_{k,c} \geq 0, \quad \forall k,\label{ob_d}\\
  &|\mathbf{f}_{l,m}|=1,~\forall l, \forall m, \label{ob_e} 
	\end{align}
\end{subequations}
where vector $\mathbf{c}=\left[C_{1,c},\dots,C_{K,c}\right]^T$ aggregates all common rate allocation variables, and $P_{\text{th}}$ denotes the maximum transmit power budget. (\ref{ob_b}), (\ref{ob_c}), and (\ref{ob_d}) enforce the transmit power limitation and common rate allocation requirement. (\ref{ob_e}) represents the unit-modulus constraint of PS elements.

Problem (\ref{linear_p}) remains elusive to optimally solve due to three technical challenges. First, the tight coupling between analog and digital beamfocusers introduces non-convexity, rendering conventional primal-dual optimization methods ineffective due to an indeterminate duality gap. Second, the expectation operator in the transmit rate expression leads to inherent non-smoothness, further exacerbating the challenges in hybrid beamfocusing design. Third, the unit-modulus constraint (\ref{ob_e}) is non-trivial and non-convex, significantly complicating the optimization process. As a result, achieving a globally optimal solution appears computationally intractable under the current framework.

\section{Proposed algorithm}\label{Section III}
This section first reformulates the objective function and constraints into tractable forms through a penalty-based approach. Building on this reformulation, we then develop a penalty-based BCD algorithm that optimizes three distinct blocks in an alternating manner.  Finally, we discuss key properties of the proposed algorithm, including its convergence and complexity.

\subsection{Problem reformulation}
To transform rate expressions in (\ref{Common_rate}) and (\ref{Private_rate}) into tractable form, we rewrite the received signal at user $k$ as:
\begin{equation}
y_{k}= \sum^K_{i=0}\hat{\mathbf{h}}^H_k\mathbf{F}\mathbf{w}_{i}x_{i}+\sum^K_{i=0}\tilde{\mathbf{h}}^H_k\mathbf{F}\mathbf{w}_{i}x_{i}+n_{k}.
\end{equation}
The second term contains multiple sources of randomness arising from the channel estimation error $\tilde{\mathbf{h}}_k$, analog beamfocuser $\mathbf{F}$, and digital beamfocuser $\mathbf{W}$. To decouple these dependencies, we adopt a worst-case noise approach by approximating this term as an independent Gaussian interference.  This conservative approximation, based on the generalized mutual information framework \cite{9894281}, provides a lower bound on the achievable rate. Consequently, we can derive the lower bound for both the common rate and private rate, as shown in (\ref{Lower_common_rate}) and (\ref{Lower_private_rate}) at the top of this page, where (a) follows Jensen's inequality.  These lower bounds are subsequently employed as the new objective function in our optimization problem. A similar approach has been adopted in\cite{10679658}.
\begin{table*}[th]
\hrule
\begin{align}
R_{k,c}&\geq \mathbb{E}_{\tilde{\mathbf{h}}_{k}}\left(\log\left(1+\frac{\left|\hat{\mathbf{h}}^H_k\mathbf{Fw}_0\right|^2}{\sum^{K}_{i=1}\left|\hat{\mathbf{h}}^H_k\mathbf{Fw}_i\right|^2 + \sum^{K}_{i=0}\left|\tilde{\mathbf{h}}^H_k\mathbf{Fw}_i\right|^2 + \sigma^2_k}\right)\right)\overset{(a)}\geq \log\left(1+\frac{\left|\hat{\mathbf{h}}^H_k\mathbf{Fw}_0\right|^2}{\sum^{K}_{i=1}\left|\hat{\mathbf{h}}^H_k\mathbf{Fw}_i\right|^2 + \sum^{K}_{i=0}\epsilon^2_k\left|\mathbf{Fw}_i\right|^2 + \sigma^2_k}\right)=\hat R_{k,c}\label{Lower_common_rate}\\
R_{k,p}&\geq \mathbb{E}_{\tilde{\mathbf{h}}_{k}}\left(\log\left(1+\frac{\left|\hat{\mathbf{h}}^H_k\mathbf{Fw}_k\right|^2}{\Delta_k\left|\mathbf{h}^H_k\mathbf{Fw}_0\right|^2+\sum^{K}_{i=1,i\neq k}\left|\hat{\mathbf{h}}^H_k\mathbf{Fw}_i\right|^2 + \sum^{K}_{i=0}\left|\tilde{\mathbf{h}}^H_k\mathbf{Fw}_i\right|^2 + \sigma^2_k}\right)\right)\notag\\&\overset{(a)}\geq \log\left(1+\frac{\left|\hat{\mathbf{h}}^H_k\mathbf{Fw}_0\right|^2}{\Delta_k\left|\mathbf{h}^H_k\mathbf{Fw}_0\right|^2+\sum^{K}_{i=1, i\neq k}\left|\hat{\mathbf{h}}^H_k\mathbf{Fw}_i\right|^2 + \sum^{K}_{i=0}\epsilon^2_k\left|\mathbf{Fw}_i\right|^2 + \sigma^2_k}\right)=\hat R_{k,p}\label{Lower_private_rate}
\end{align}
\hrule
\end{table*}

To attack the coupling between analog and digital beamfocusers, we introduce an auxiliary matrix $\mathbf{P}=\left[\mathbf{p}_0,\mathbf{p}_1,\dots, \mathbf{p}_K\right]\in\mathbb{C}^{N\times (K+1)}$, where $\mathbf{p}_k=\mathbf{Fw}_k$ for $\forall k\in\tilde{\mathcal{K}}$. Plugging $\mathbf{p}_k=\mathbf{Fw}_k$ into (\ref{Lower_common_rate}) and (\ref{Lower_private_rate}), we have
\begin{align}
\hat R_{k,c}&=\log\left(1+\frac{\left|\hat{\mathbf{h}}^H_k\mathbf{p}_0\right|^2}{\sum^{K}_{i=1}\left|\hat{\mathbf{h}}^H_k\mathbf{p}_i\right|^2 + \hat\sigma^2_k}\right)\label{Recast_common_rate},\\
\hat R_{k,p}&=\log\left(1+\frac{\left|\hat{\mathbf{h}}^H_k\mathbf{p}_k\right|^2}{\sum^{K}_{\underset{i\neq k}{i=1}}\Delta_{k,i}\left|\hat{\mathbf{h}}^H_k\mathbf{p}_i\right|^2 + \hat\sigma^2_k}\right),
\label{Recast_private_rate}
\end{align}
where $\hat \sigma^2_k =\sum^{K}_{i=0}\epsilon^2_k\left|\mathbf{p}_i\right|^2 + \sigma^2_k $ and
\begin{align}
\Delta_{k,i}&=\begin{cases}
1, \quad\qquad&\mbox{if $i\neq 0$ },\\
\Delta_k,  &\mbox{if $i=0$ }.
\end{cases}
\end{align}
Meanwhile, to overcome non-smoothness, we remove the minimum operator by introducing a non-negative auxiliary variable $\hat R$. This allows us to reformulate problem (\ref{linear_p}) as 
\begin{subequations}\label{linear_p2}
	\begin{align}
&\max_{\mathbf{F},\mathbf{W},\mathbf{P},\mathbf{c},\hat R} \hat R,\label{ob_a2}\\
	\text{s.t.}~
	&\mathbf{P}=\mathbf{F}\mathbf{W},\label{ob_b2}\\
 &||\mathbf{P}||^2_F\leq P_{\text{th}},\label{ob_c2}\\
 &C_{k,c}+\hat R_{k,p}\geq \hat R,~\forall k,\label{ob_d2}\\
 &\sum^K_{k=1}C_{k,c}\geq \hat R_{k,c},~\forall k,\label{ob_e2}\\
 &\mbox{ (\ref{ob_d}),~ (\ref{ob_e})}.\label{ob_f2}
	\end{align}
\end{subequations}
However, the equality constraint (\ref{ob_b2}) complicates the direct hybrid beamfocusing design. To attack this difficulty, we incorporate the equality constraint into the objective function as the penalty term. This reformulation yields the new optimization problem:
\begin{subequations}\label{linear_p3}
	\begin{align}
&\max_{\mathbf{F},\mathbf{W},\mathbf{P},\mathbf{c},\hat R} \hat R -\frac{1}{\rho}||\mathbf{P}-\mathbf{FW}||^2_F,\label{ob_a3}\\
	\text{s.t.}~
 &\mbox{ (\ref{ob_d}),~(\ref{ob_e}),~(\ref{ob_c2}),~(\ref{ob_d2}),~(\ref{ob_e2})},\label{ob_b3}
	\end{align}
\end{subequations}
where $\rho>0$ is the penalty parameter. Note that when $\rho\to 0$, the equality constraint $\mathbf{P}=\mathbf{FW}$ will be exactly satisfied. However, to effectively maximize the transmission rate, we implement a continuation strategy: the penalty parameter is initialized with a sufficiently large value and then gradually decreased. This naturally leads to a double-loop algorithmic framework, where the inner loop solves problem (\ref{linear_p3}) for a fixed $\rho$ while the outer loop updates the penalty factor. We next focus on solving problem (\ref{linear_p3})  for a given penalty factor.

\subsection{Proposed solution for solving problem (\ref{linear_p3}) }
Although the penalty factor is fixed, problem (\ref{linear_p3}) remains challenging to solve due to the coupled variables. To attack this issue, we decompose optimization variables into three blocks, namely, $\big\{\mathbf{P},\mathbf{c},\hat R\big\}$, $\mathbf{F}$, and $\mathbf{W}$. Then, we employ the BCD algorithmic framework to alternately optimize each block, with the detailed procedure provided next.

\subsubsection{Subproblem w.r.t. $\big\{\mathbf{P},\mathbf{c},\hat R\big\}$ } With the known $\mathbf{F}$ and $\mathbf{W}$, problem (\ref{linear_p3}) reduces to
\begin{subequations}\label{linear_p4}
	\begin{align}
&\max_{\mathbf{P},\mathbf{c},\hat R} \hat R -\frac{1}{\rho}||\mathbf{P}-\mathbf{FW}||^2_F,\label{ob_a4}\\
	\text{s.t.}~
  &\mbox{\mbox{ (\ref{ob_d}),~(\ref{ob_c2}),~(\ref{ob_d2}),~(\ref{ob_e2})}}.\label{ob_b4}
	\end{align}
\end{subequations}
To efficiently solve this subproblem, we use surrogate optimization\cite{lange2000optimization}, a technique that replaces the complex objective function with a tractable surrogate. A critical prerequisite is that the surrogate must closely approximate the original function while remaining computationally tractable. Herein, we build lower-bounded concave quadratic surrogates to approximate logarithmic transmit rates. Specifically, the concave quadratic surrogates for $R_{k,c}$ and $R_{k,p}$ are respectively given by
\begin{subequations}\label{Overall_SCA}
\begin{align}
&f_{k,c}\left(\mathbf{P}\right)= \sum_{j=0}^{K}\mathbf{p}^H_{j}\mathbf{x}_{k,c}\mathbf{p}_{j}+2\text{Re}\left(\mathbf{y}_{k,c}\mathbf{p}_{0}\right)+ z_{k,c},
\label{SCA_1}\\
&f_{k,p}\left(\mathbf{P}\right)= \sum_{j=0}^{K}\Delta_{k,j}\mathbf{p}^H_{j}\mathbf{x}_{k,p}\mathbf{p}_{j}+2\text{Re}\left(\mathbf{y}_{k,p}\mathbf{p}_{k}\right) + z_{k,p}
\label{SCA_2}
\end{align}
\end{subequations}
where 
\begin{align}
\mathbf{x}_{k,\tau}&=-\frac{1}{\ln 2}\hat{\mathbf{h}}_k\tilde{\mathbf{u}}_{k,\tau} \left(\tilde v_{k,\tau}\right)^{-1}\tilde{\mathbf{u}}^H_{k,\tau}\hat{\mathbf{h}}^H_k,\notag\\
\mathbf{y}_{k,\tau}&=\frac{1}{\ln 2}\left(\tilde v_{k,\tau}\right)^{-1}\tilde{\mathbf{u}}^H_{k,\tau}\hat{\mathbf{h}}^H_k,\\
z_{k,\tau}&=\frac{1}{\ln 2}-\frac{1}{\ln 2}\left(\tilde v_{k,\tau}\right)^{-1}\left(\tilde{\sigma}^2_{k}\tilde{\mathbf{u}}^H_{k,\tau}\tilde{\mathbf{u}}_{k,\tau}+1\right)-\log \tilde{v}_{k,\tau},\notag
\end{align}
with $\forall\tau\in\{c,p\}$, $\tilde{\sigma}^2_{k}=\sum^{K}_{i=0}\epsilon^2_k\left|\tilde{\mathbf{p}}_i\right|^2 + \sigma^2_k$, and 
\begin{align}\label{Auxiliary_1}
&\tilde{\mathbf{u}}_{k,c}=\left(\sum_{j=0}^{K}\hat{\mathbf{h}}^H_k\tilde{\mathbf{p}}_{j}\tilde{\mathbf{p}}^H_{j}\hat{\mathbf{h}}_k+\hat{\sigma}^2_{k}\right)^{-1}\hat{\mathbf{h}}^H_k\tilde{\mathbf{p}}_{0},\notag\\
&\tilde v_{k,c}=1-\tilde{\mathbf{u}}^H_{k,c}\hat{\mathbf{h}}^H_k\tilde{\mathbf{p}}_{0},\\
&\tilde{\mathbf{u}}_{k,p}=\left(\sum_{j=0}^{K}\Delta_{k,j}\hat{\mathbf{h}}^H_k\tilde{\mathbf{p}}_{j}\tilde{\mathbf{p}}^H_{j}\hat{\mathbf{h}}_k+\hat{\sigma}^2_{k}\right)^{-1}\hat{\mathbf{h}}^H_k\tilde{\mathbf{p}}_{k},\notag\\
&\tilde v_{k,p}=1-\tilde{\mathbf{u}}^H_{k,p}\hat{\mathbf{h}}^H_k\tilde{\mathbf{p}}_{k}.\notag
\end{align}
$\tilde{\mathbf{p}}_{k}$ in equation (\ref{Auxiliary_1}) denotes the expansion point of beamfocusing vector $\mathbf{p}_{k}$ for $\forall k\in\tilde{\mathcal{K}}$. Our defined surrogates possess minorization property and gradient consistency, thereby guaranteeing good convergence as established in \cite{mairal2013optimization}.

{\bf{Claim 1:}} The constructed surrogates $f_{k,\tau}\left(\mathbf{P}\right)$ meet:
\begin{enumerate}
    \item \emph{Minorization property:} $f_{k,\tau}\left(\mathbf{P}\right)$ strictly lower-bounds the transmit rate, i.e., $\hat R_{k,\tau}\geq f_{k,\tau}\left(\mathbf{P}\right)$ with equality holds when $\mathbf{p}_{k}=\tilde{\mathbf{p}}_{k}$ for $\forall k$. 
    \item \emph{Gradient consistency:} $f_{k,\tau}\left(\mathbf{P}\right)$ maintains first-order consistency with the transmit rate at the expansion point $\tilde{\mathbf{p}}_{k}$, i.e.,
    \begin{equation}
\frac{\partial{f_{k,\tau}\left(\mathbf{P}\right)}}{\partial{\mathbf{p}_{k}}}\Bigg|_{\mathbf{p}_{k} = \tilde{\mathbf{p}}_{k}}=\frac{\partial{\hat R_{k,\tau}}}{\partial{\mathbf{p}_{k}}}\Bigg|_{\mathbf{p}_{k} = \tilde{\mathbf{p}}_{k}}.
    \end{equation}
\end{enumerate}
\emph{Proof:} This proof shares a similar methodology with \cite{8976409}. Please see \cite{8976409} for detailed proof.\hfill \QEDclosed

Based on the constructed surrogates (\ref{Overall_SCA}), problem (\ref{linear_p4}) can be reformulated as
\begin{subequations}\label{linear_p5}
	\begin{align}
&\max_{\mathbf{P},\mathbf{c},\hat R} \hat R -\frac{1}{\rho}||\mathbf{P}-\mathbf{FW}||^2_F,\label{ob_a5}\\
	\text{s.t.}~&\sum_{k=1}^{K}C_{k,c} \leq f_{k,c}\left(\mathbf{P}\right),~\forall k,\label{ob_b5}\\
    &C_{k,c} +f_{k,p}\left(\mathbf{P}\right)\geq \hat R,~\forall k,\label{ob_c5}\\
  &\mbox{ (\ref{ob_d}), (\ref{ob_c2})}.\label{ob_d5}
	\end{align}
\end{subequations}
Problem (\ref{linear_p5}) becomes convex when the expansion point $\tilde{\mathbf{p}}_{k}$ is fixed,  enabling efficient solution via standard off-the-shelf solvers. The propsoed iterative algorithm is summarized in Algorithm \ref{Alg.1}.
\begin{algorithm}[t]
	\caption{Iterative algorithm for solving (\ref{linear_p4})}
	\begin{algorithmic}[1]\label{Alg.1}
		\STATE Initialize $\mathbf{p}_{k}$ for $\forall k$. 
		\REPEAT
        \STATE Update $\tilde{\mathbf{p}}_{k}=\mathbf{p}_{k}$.
        \STATE Update surrogate $f_{k,\tau}\left(\mathbf{P}\right)$ based on equation (\ref{Overall_SCA}).
		\STATE  Solving problem (\ref{linear_p5}) to obtain optimal $\mathbf{p}_{k}$.
		\UNTIL{the increment of the objective value of problem  (\ref{linear_p4}) falls below a threshold.}	
		\STATE Output the optimized $\mathbf{P}$.
	\end{algorithmic}
\end{algorithm}
\subsubsection{Subproblem w.r.t. $\mathbf{F}$ } The variables $\mathbf{F}$ only appear in the last term of the objective function, leading to the following subproblem:
\begin{subequations}\label{linear_s1}
	\begin{align}
&\min_{\mathbf{F}} ||\mathbf{P}-\mathbf{F}\mathbf{W}||^2_F,\label{ob_sa1}\\
	\text{s.t.}~
  &\mbox{ (\ref{ob_e})}.\label{ob_sb1}
	\end{align}
\end{subequations}
Since the analog beamfocuser $\mathbf{F}$ is a block-diagonal matrix, we can rewrite the objective function as follows:
\begin{align}\label{New_obj_2}
&||\mathbf{P}-\mathbf{F}\mathbf{W}||^2_F=\sum^{L}_{l=1}||\hat{\mathbf{P}}_{l}-\mathbf{f}_{l}\hat{\mathbf{w}}_{l}||^2_F\notag\\
=&\hat\eta-\sum^{L}_{l=1}\sum^{M}_{m=1}2\text{Re}\left(\mathbf{\psi}^H_{l,m}\mathbf{f}_{l,m}\right),
\end{align}
where 
$\hat{\mathbf{P}}_{l}=\mathbf{P}\big((l-1)M+1:lM,:\big)$,
$\hat{\mathbf{w}}_{l} = \mathbf{W}(l,:)$, $\hat \eta =\sum^{L}_{l=1}\left(M\hat{\mathbf{w}}_{l}\hat{\mathbf{w}}^H_{l}+\text{Tr}\left(\hat{\mathbf{P}}_{l}\hat{\mathbf{P}}^H_{l}\right)\right)$,$\mathbf{\Psi}_{l}=\hat{\mathbf{P}}_{l}\hat{\mathbf{w}}^H_{l}$, and $\mathbf{\psi}_{l,m}$ is the $m$-th element of $\mathbf{\Psi}_{l}$.
Based on equation (\ref{New_obj_2}), we observe that problem (\ref{linear_s1}) can be decomposed into $ML$ independent subproblem. Specifically, the subproblem regarding the optimization of $\mathbf{f}_{l,m}$  is
\begin{subequations}\label{linear_e2}
	\begin{align}
&\max_{\mathbf{f}_{l,m}}\text{Re}\left(\mathbf{\psi}^H_{l,m}\mathbf{f}_{l,m}\right),\label{ob_ea2}\\
	\text{s.t.}~
  &|\mathbf{f}_{l,m}|=1.\label{ob_eb2}
	\end{align}
\end{subequations}
The optimal $\mathbf{f}^*_{l,m}$ is given by
\begin{equation}
\mathbf{f}^*_{l,m}=e^{-j\angle\mathbf{\psi}^H_{l,m}}.
\label{PS_beamfocusing_2}
\end{equation}

\subsubsection{Subproblem w.r.t. $\mathbf{W}$ } The optimal solution of $\mathbf{W}$ can be obtained by minimizing $||\mathbf{P}-\mathbf{F}\mathbf{W}||^2_F$ without any constraint, which is given by
\begin{align}\label{linear_p6}
\mathbf{W}^*=&\arg\min_{\mathbf{W} } ||\mathbf{P}-\mathbf{FW}||^2_F\notag\\
=&\left(\mathbf{F}^H\mathbf{F}\right)^{-1}\mathbf{F}^H\mathbf{P}.
\end{align}

\subsection{Overall algorithm and properties analysis}
The proposed penalty-based BCD algorithm to solve the problem (\ref{linear_p3}) is summarized in Algorithm~\ref{Alg.2}. Its convergence and complexity are analyzed below:
\begin{itemize}
\item  \emph{Convergence}: From an arbitrary initial point, Algorithm~\ref{Alg.2} produces an non-decreasing sequence of objective values during its inner loop iterations. Since the achievable transmit rate is lower-bounded, the inner loop iteration converges to a stationary point. Furthermore, the convergence of outer-loop penalty-based approach has been proven in\cite{9120361}. We thus conclude Algorithm~\ref{Alg.2} attains convergence within finite iterations.

\item \emph{Complexity}: Algorithm~\ref{Alg.1} has complexity  $\mathcal O\left(N_v^{3.5}\right)$, where $N_v=NK+K+N$ is the number of optimization variables.   The updates of $\mathbf{F}$ and $\mathbf{W}$ have the complexity $\mathcal O\left(M^2L(K+1)\right)$ and $\mathcal O\left(NL\max(L,K+1)\right)$ respectively. Consequently, the per-iteration complexity of Algorithm~\ref{Alg.2} is in order of $\mathcal O\left(\delta N_v^{3.5} + M^2L(K+1)+ NL\max(L,K+1)\right)$, where $\delta$ denotes the iteration number until Algorithm~\ref{Alg.1} converges.
\end{itemize}

\begin{algorithm}[t]
	\caption{Penalty-based BCD algorithm for solving (\ref{linear_p3})}
	\begin{algorithmic}[1]\label{Alg.2}
		\STATE Initialize $\mathbf{F}$ and $\mathbf{W}$.
        \REPEAT
		\REPEAT
		\STATE  Update $\mathbf{P}$ by calling Algorithm~\ref{Alg.1}.
        \STATE Update $\mathbf{F}$ based on equation (\ref{PS_beamfocusing_2}).
        \STATE Update $\mathbf{W}$ based on equation (\ref{linear_p6}).
		\UNTIL { the increment of the objective value of problem (\ref{linear_p3}) falls below a threshold. }
        \STATE Update penalty factor $\rho=\alpha  \rho$.
        \UNTIL{the penalty value falls below a threshold. }
		\STATE Output the optimized max-min rate $\hat R$.
	\end{algorithmic}
\end{algorithm}

\section{Low-complexity algorithm design}\label{Section IV}
Although the penalty-based BCD algorithm can converge to a stationary point, it relies on double-loop iterations to optimize the introduced fully digital beamfocuser, analog beamfocuser, and digital beamfocuser. In NFC with ELAAs, the high dimensionality of these beamfocusers leads to substantial computational complexity. To mitigate this issue, we propose a low-complexity algorithm that decouples the design of analog and digital beamfocusers into two separate stages.

\subsection{Heuristic analog beamfocuser design}
Similar to\cite{10559261,10587118}, we consider designing the analog beamfocuser to maximize the array gain for each user. To ensure fairness across all users, the  optimization problem for the analog beamfocuser can be formulated as
\begin{subequations}\label{linear_l1}
	\begin{align}
&\max_{\mathbf{F}}\min_{\forall k} ||\mathbf{a}^H_k\mathbf{F}||^2,\label{ob_la}\\
	\text{s.t.}~
 &|\mathbf{f}_{l,m}|=1,~\forall l, \forall m.\label{ob_lb} 
	\end{align}
\end{subequations}
The solution to problem (\ref{linear_l1}) presents significant challenges due to the unit-modulus constraint and the block-diagonal structure of $\mathbf{F}$. Prior works randomly allocate an equal number of RF chains to each user under a fully-connected PS architecture \cite{10559261,10587118}. However, our sub-connected PS framework enforces a block-diagonal structure for the analog beamfocuser. Under this architecture, random RF chain assignment may lead to substantial performance degradation. To address this challenge, we propose an enhanced RF chain assignment by incorporating random allocation and swap operations. 

The proposed analog beamfocuser design process comprises two steps. In the first step, each user is randomly allocated $\left\lfloor \frac{L}{K}\right\rfloor $ RF chains, then the remaining $L-\left\lfloor \frac{L}{K}\right\rfloor K$ RF chains are randomly allocated to $L-\left\lfloor \frac{L}{K}\right\rfloor K$ distinct users. Let $\Phi$ is the RF chain allocation results and $k=\Phi\left(l\right)$ indicates that $l$-th RF chain serves the $k$-th user.  Assuming $k=\Phi\left(l\right)$, the optimal analog beamfocuser $\mathbf{f}_l$ is given by
\begin{align}
\mathbf{f}^*_l=&\arg\max_{\mathbf{f}_l } ||\mathbf{a}_k\mathbf{f}_l||^2=\mathbf{a}_k\left(M(l-1)+1:Ml\right).
\label{uti_k}
\end{align}
Integrating the allocation result $\Phi$ and equation (\ref{uti_k}), we can obtain the initial analog beamfocuser matrix $\mathbf{F}$.

The second step aims to maximize the minimum array gain among all users. To this end, we enable swap operations among two RF chains to switch their served users while keeping other results unchanged. A swapping operation is 
\begin{align}
\Phi^{l'}_l=\left\{\Phi\backslash\left\{\left(k,l\right),\left(k',l'\right)\right\}\cup\left\{\left(k,l'\right),\left(k',l\right)\right\}\right\}.
\label{swap}
\end{align}
where $k=\Phi\left(l\right)$ and $k'=\Phi\left(l'\right)$. A swap operation is permitted only when it constitutes a swap-blocking pair, defined by the following criterion: $\left(l,l'\right)$ is a swap-blocking pair if and only if the minimum array gain among all users can be improved, i.e., $\min_{\forall k} ||\mathbf{a}^H_k\mathbf{F}'||^2 \geq \min_{\forall k} ||\mathbf{a}^H_k\mathbf{F}||^2$, where $\mathbf{F}'$ denotes the updated analog beamfocuser matrix after the swap operation. The RF chain allocation result $\Phi$ can output a stable result when there is no swap-blocking pair.

\subsection{Digital beamfocuser design}
Given the known analog beamfocuser designed in the previous subsection, the rates of decoding common and private streams at $k$-th user are respectively given by
\begin{align}
\hat R_{k,c}&=\log\left(1+\frac{\left|\bar{\mathbf{h}}^H_k\mathbf{w}_0\right|^2}{\sum^{K}_{i=1}\left|\bar{\mathbf{h}}^H_k\mathbf{w}_i\right|^2 + \bar\sigma^2_k}\right)\label{Recast_common_rate}\\
\hat R_{k,p}&=\log\left(1+\frac{\left|\bar{\mathbf{h}}^H_k\mathbf{w}_k\right|^2}{\sum^{K}_{i=0,i\neq k}\Delta_{k,i}\left|\bar{\mathbf{h}}^H_k\mathbf{w}_i\right|^2 + \bar\sigma^2_k}\right)
\label{Recast_private_rate}
\end{align}
where $\bar \sigma^2_k =\sum^{K}_{i=0}\epsilon^2_k\left|\mathbf{Fw}_i\right|^2 + \sigma^2_k $ and $\bar{\mathbf{h}}_k=\mathbf{F}^H\hat{\mathbf{h}}_k$ denotes the equivalent channel, which has a low dimensionality of $L$. Then, the digital beamfocuser optimization can be reformulated as
\begin{subequations}\label{linear_p9}
	\begin{align}
&\max_{\mathbf{W},\mathbf{c},\hat R } \hat R \label{ob_a9}\\
	\text{s.t.}~
 &\mbox{ (\ref{ob_b}),~(\ref{ob_d}),~(\ref{ob_d2}),~(\ref{ob_e2})}.\label{ob_b9} 
	\end{align}
\end{subequations}
This problem can be efficiently solved by the proposed Algorithm~\ref{Alg.1}, where $\mathbf{p}_j$, $\tilde{\mathbf{p}}_j$, $\hat{\mathbf{h}}_k$, and $\hat{\sigma}^2_k$ in the quadratic surrogates are replaced by $\mathbf{w}_j$, $\tilde{\mathbf{w}}_j$, $\bar{\mathbf{h}}_k$, and $\bar{\sigma}^2_k$, respectively. 

\subsection{Overall low-complexity algorithm and properties analysis}
\begin{algorithm}[t]
	\caption{Low-complexity algorithm for solving (\ref{linear_p})}
	\begin{algorithmic}[1]\label{Alg.3}
    \STATE {\bf{State 1:} Heuristic analog beamfocuser design}
		\STATE Initialize the RF chain allocation result $\Phi$.
        \STATE Obtain the initial analog beamfocuser matrix $\mathbf{F}$.
        \REPEAT
        \STATE $l$-th RF chains search for another RF chain for $\forall l$ to form a swap-blocking pair.
        \IF{$(l,l')$ is a swap-blocking pair}
		\STATE Update the allocation result $\Phi$ and analog beamfocuser matrix $\mathbf{F}$.
		\ELSE
		\STATE Keep the current allocation state.
		\ENDIF	
		\UNTIL{There is no swap-blocking pair.}
        \STATE {\bf{State 2: Digital beamfocuser optimization}}
        \REPEAT
        \STATE Update $\tilde{\mathbf{w}}_{k}=\mathbf{w}_{k}$.
        \STATE Construct surrogates according to equation (\ref{Overall_SCA}).
		\STATE  Solving problem (\ref{linear_p9}) to obtain optimal $\mathbf{w}_{k}$.
		\UNTIL{the increment of the objective value of problem  (\ref{linear_p9}) falls below a threshold.}	
		\STATE Output the optimized max-min rate $\hat R$.
	\end{algorithmic}
\end{algorithm} 
Based on the above insights, the overall low-complexity algorithm for solving problem (\ref{linear_p}) is summarized as Algorithm~\ref{Alg.3}. Its critical properties are discussed below.
\begin{itemize}
\item  \emph{Convergence}: Given an arbitrary initial allocation result, the minimum array gain increases after each swap operation. Considering the array gain cannot increase infinitely, we thus deduce that the heuristic analog beamfocuser design yields a deterministic analog beamfocusing matrix $\mathbf{F}$. As discussed in Section III.C, stage 2 in Algorithm~\ref{Alg.3} converges within finite iterations.

\item \emph{Complexity}: The complexity for updating the analog beamfocusing matrix and computing the array gain are $\mathcal{O}\left(L\right)$ and  $\mathcal{O}\left(NKL\right)$, respectively. Thus, the complexity in stage 1 is $\mathcal{O}\left(L+NKL\right)$. In stage 2, the complexity is $O\left(\delta (LK+L+K)^{3.5}\right)$. Consequently, the overall complexity of Algorithm~\ref{Alg.3} is $\mathcal O\left(L+NKL+\delta_1 (LK+L+K)^{3.5}\right)$, where $\delta_1$ denotes the iteration number until the stage 2 converges. Furthermore, Algorithm~\ref{Alg.3} eliminates double loop iteration, which is more efficient than Algorithm~\ref{Alg.2}.
\end{itemize}

\section{Simulation results}\label{Section V}
This section provides the numerical results to evaluate our proposed transmit scheme and algorithms. Unless stated otherwise, the simulation settings are as follows: a BS equipped with $N=128$ antennas and $L=8$ RF chains operates at a frequency of $f_c=30$~GHz. The inter-element spacing is the half-wavelength. $K=4$ users are randomly generated within the distance from $10$~m to $20$~m. The maximum transmit power and background noise power are $P_{\text{th}}=20$~dBm and $\sigma^2_k=-84$~dBm. The channel estimation error variance and SIC imperfection factor are set to $\epsilon^2_k=0.005||\hat{\mathbf{h}}_k||^2$ and $\Delta_k=0.05$, respectively. The penalty factors are set to $\rho=10^2$, with a reduction factor of $\alpha=0.5$. These parameter settings align with prior works\cite{10579914,10587118}.

Across $100$ independent channel realizations, our proposed transmit scheme and algorithms (labeled {\bf{RSMA-SHB}} and {\bf{RSMA-SHB-Low}}) are compared against three benchmarks for a thorough performance evaluation. They are described below.

\begin{itemize}
\item {\bf{RSMA-FD}}: Each antenna is connected to a dedicated RF chain, enabling full digital beamfocusing capability. This benchmark establishes the theoretical performance upper bound for our proposed sub-connected HAD architecture.
\item {\bf{SDMA-SHB}}: This benchmark leverages the spatial beamfocusing exclusively to mitigate interference. Specifically, it adopts the same sub-connected HAD architecture, but employs conventional user-specific stream encoding, e.g., $\mathbf{w}_{0}=\mathbf{0}$. At the receiver end, each user directly decodes its desired stream by treating all interference as noise. 
\item {\bf{RSMA-SHB-far}}: This baseline implements conventional far-field beamforming using the array response vector:
\begin{equation}
\mathbf{a}_{\text{far}}\left(\theta_k\right)= \left[e^{j\frac{2\pi}{\lambda}\frac{1-N}{2}d\sin\theta_k},\dots,e^{j\frac{2\pi}{\lambda}\frac{N-1}{2}d\sin\theta_k}\right]^T.
\label{Far-Channel}
\end{equation}
In addition, all other parameters remain unchanged to ensure fair comparison.
\end{itemize}

\begin{figure}[tbp]
	\centering
	\includegraphics[scale=0.5]{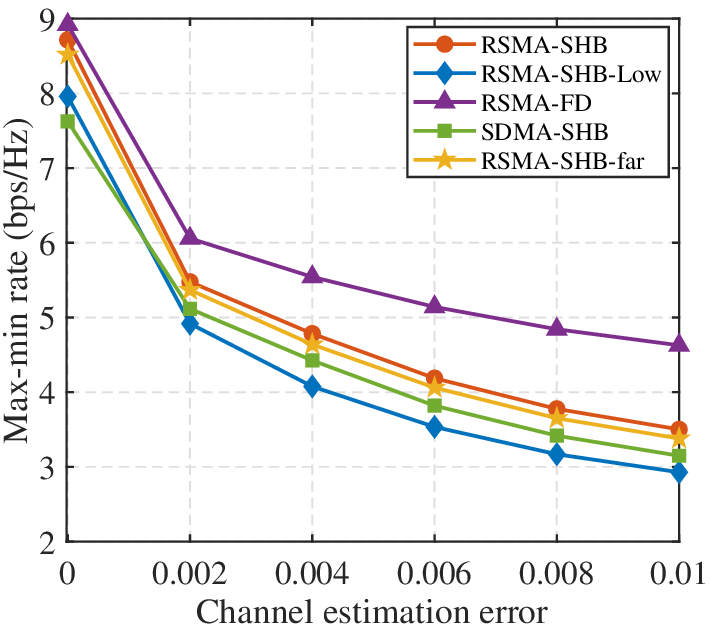}
	\caption{Max-min rate versus channel estimation error.}
	\label{error}
\end{figure}
Fig.~\ref{error} illustrates the max-min rate against channel estimation error under perfect SIC conditions, revealing three key insights. First, our proposed transmit scheme consistently surpasses both conventional far-field beamforming and SDMA-enabled beamfocusing. The superiority suggests that while beamfocusing partially suppresses MUI, it cannot fully eliminate such interference even with perfect CSI. Second, the proposed low-complexity algorithm provides a comparison of performance relative to the penalty-based BCD method while avoiding computationally expensive double-loop iterations, underscoring its practical efficiency. Third, as channel estimation error increases, the performance gap between the sub-connected HAD and full digital beamfocusing architectures widens. Despite this, our approach reduces RF chain requirements by 16 times. Additionally, our scheme maintains near-optimal full digital beamfocusing performance under perfect CSI conditions.

\begin{figure}[tbp]
	\centering
	\includegraphics[scale=0.5]{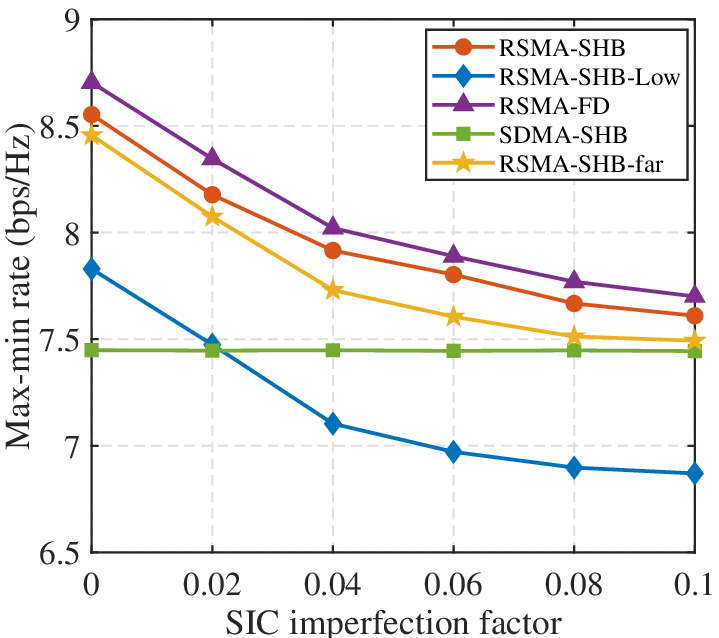}
	\caption{Max-min rate versus SIC imperfection factor.}
	\label{SIC}
\end{figure}
Fig.~\ref{SIC} presents the relationship between the max-min rate and the imperfect SIC factor under perfect channel information. As anticipated, the achievable rate of the SDMA-based transmit scheme remains constant regardless of the SIC error level. This behavior occurs because SDMA directly decodes the desired stream, rendering its performance independent of SIC imperfections. In contrast, all other schemes exhibit a progressive degradation in the max-min rate as the SIC imperfection factor increases, primarily due to the reduced decoding rate of private streams at higher imperfection levels. However, we observe that RSMA always surpasses SDMA in performance, further highlighting two findings: 1) near-field beamfocusing alone is insufficient to fully suppress MUI, and 2) RSMA provides adaptive interference management in NFC, even under imperfect SIC conditions.

\begin{figure}[tbp]
	\centering
	\includegraphics[scale=0.5]{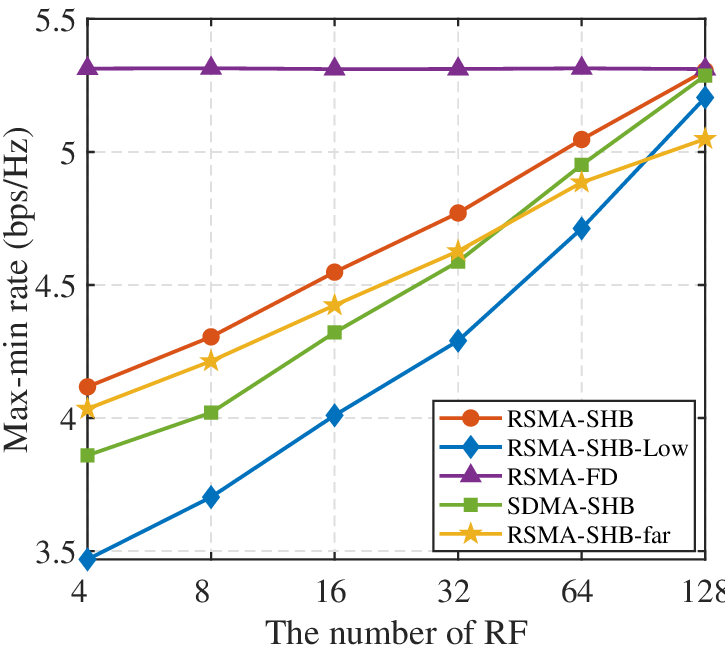}
	\caption{Max-min rate versus the number of RFs.}
	\label{RF}
\end{figure}
Fig.~\ref{RF} simulates the max-min rate versus the number of RF chains in perfect CSI and SIC conditions, indicating three interesting observations over competing benchmarks. First, as the number of RF chains increases, the performance gap between near-field beamfocusing and far-field beamforming widens. This occurs because near-field beamfocusing harnesses additional RF chains for sharper energy focusing, thereby suppressing interference more effectively. Second, near-field beamfocusing alone is capable of nearly eliminating interference under full digital architectures, rendering advanced anti-interference strategies unnecessary. Third, with an increasing number of RF chains, our proposed low-complexity algorithm delivers near-optimal performance, narrowing the gap with the high-complexity BCD algorithm. For example, at $L=128$, the performance degradation remains below $0.1$~bps/Hz.

\begin{figure}[tbp]
	\centering
	\includegraphics[scale=0.5]{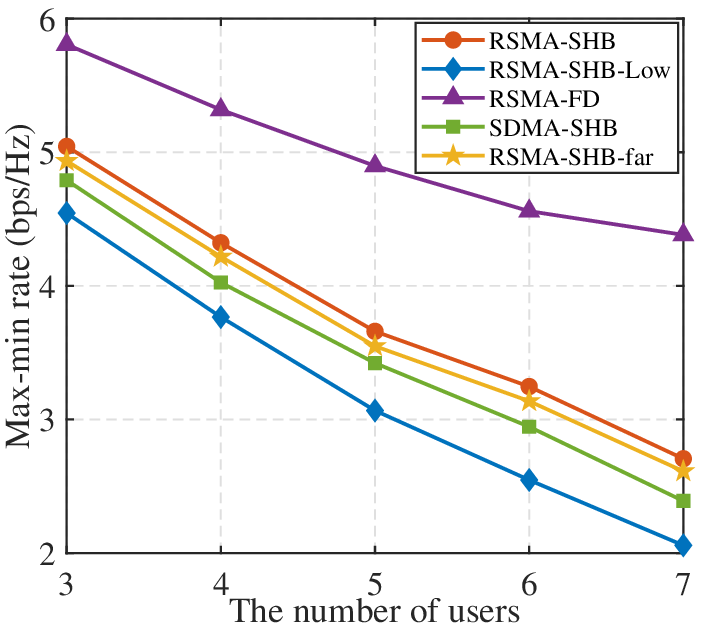}
	\caption{Max-min rate versus the number of users.}
	\label{user}
\end{figure}
Fig.~\ref{user} describes the max-min rate versus the number of users in imperfect CSI and SIC. The simulation result reveals that all approaches experience a degradation of the transmit rate as more users are scheduled, primarily due to intensified interference. Notably, the performance gap between full digital and sub-connected HAD beamfocusing architectures becomes more noticeable with increasing user numbers. This divergence arises because the former provides superior user discrimination and interference suppression, whereas the latter, with limited RF chains, cannot efficiently focus beam energy on designated locations. Despite this, compared to conventional beamfocusing, our transmit scheme delivers $0.3$~bps/Hz rate improvement, highlighting its enhanced interference management.

\begin{figure}[tbp]
	\centering
	\includegraphics[scale=0.5]{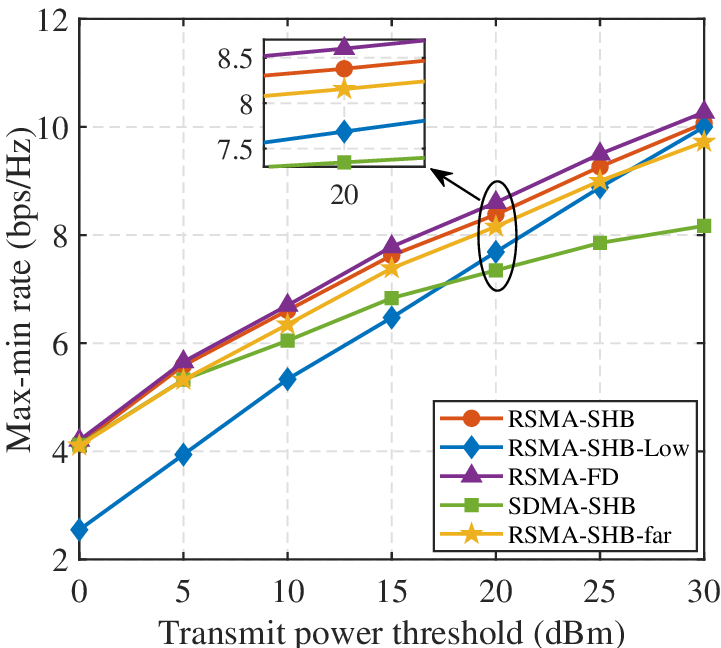}
	\caption{Max-min rate versus transmit power threshold.}
	\label{power}
\end{figure}
Fig.~\ref{power} simulates the max-min rate versus transmit power under perfect CSI and SIC conditions. We observe that the sub-connected HAD architecture achieves performance comparable to the full-digital architecture when perfect CSI and SIC are available. Moreover, RSMA is always superior to SDMA across all transmit power levels, with the performance gap widening as power increases, reaching $2$~bps/Hz at $P_{\text{th}}=30$~dBm. Additionally, SDMA exhibits a saturation trend, highlighting its limited flexibility in MUI management. Similar to Fig.~\ref{RF}, our proposed low-complexity algorithm delivers near-optimal performance, nearly matching the high-complexity BCD algorithm.

\begin{figure}[tbp]
	\centering
	\subfigure{
		\begin{minipage}[tbp]{0.45\linewidth}
			\centering\includegraphics[width = 1.53in]{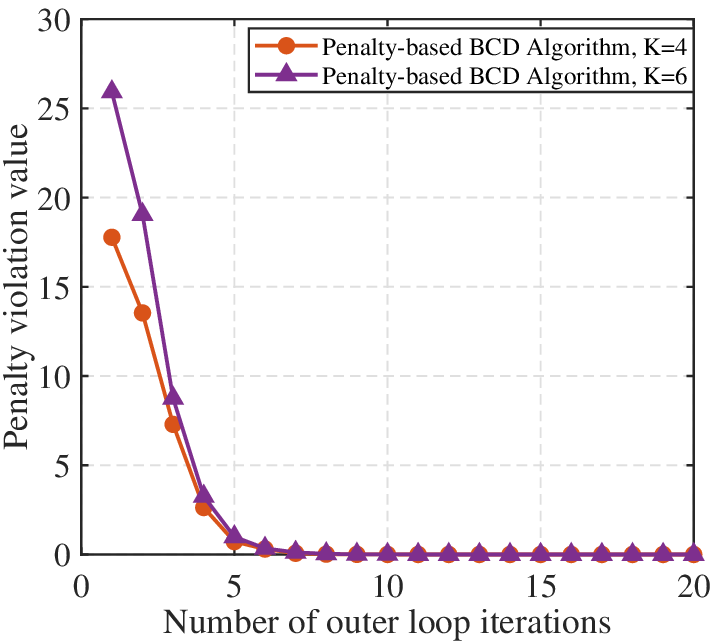}
		\end{minipage}
	}
	\subfigure{
		\begin{minipage}[tbp]{0.48\linewidth}
			\centering\includegraphics[width = 1.53in]{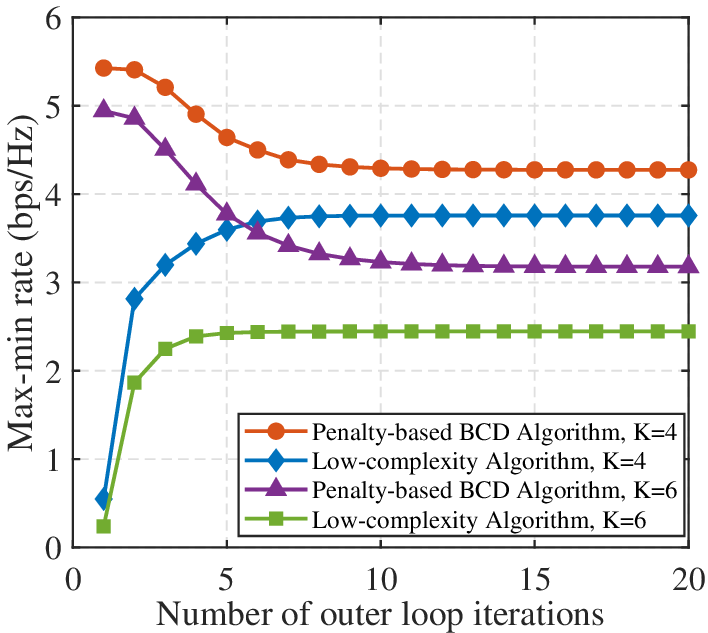}
		\end{minipage}
	}
	\centering
	\caption{Convergence behavior of the proposed algorithms.}
	\label{convergence}
\end{figure}
Fig.~\ref{convergence} examines the convergence performance of the proposed algorithms, comparing the penalty violation value (left) and the max-min rate (right) against outer-loop iterations. The results show that the penalty violation metric $||\mathbf{P}-\mathbf{FW}||^2_F$ converges rapidly, strictly enforcing the equality constraint $\mathbf{P}=\mathbf{FW}$. The penalty-based BCD algorithm presents monotonic decreasing convergence due to progressively tighter feasible solution spaces with each iteration. In comparison, our low-complexity algorithm directly maximizes the achievable rate, exhibiting monotonic performance improvement.

\section{Conclusion}\label{Section VI}
This paper considers an RSMA-enabled transmit scheme for NFC under imperfect CSI and SIC, where a sub-connected HAD architecture is adopted to reduce hardware requirements. The analog beamfocuser, digital beamfocuser, and common rate allocation are jointly optimized to maximize the minimum rate. To solve the formulated non-convex problem, we develop a penalty-based BCD algorithm, which derives closed-form solutions for the optimal analog and digital beamfocusers. Furthermore, a low-complexity hybrid beamfocusing design algorithm is proposed to avoid double-loop iterations, reducing computational complexity. Simulation results reveal that near-field beamfocusing alone is insufficient to fully mitigate MUI even under perfect CSI, underscoring the necessity of advanced interference management techniques for NFC. Additionally, near-field beamfocusing exhibits considerable performance gains over far-field beamforming. These insights provide valuable guidance for the design of efficient transmission schemes in practical NFC scenarios.

	\ifCLASSOPTIONcaptionsoff
	\newpage
	\fi
	
	\bibliographystyle{IEEEtran}
	\bibliography{references}
	
\end{document}